\documentclass[aps,prb,showpacs,twocolumn,superscriptaddress,preprintnumbers,amsmath,amssymb]{revtex4}
\usepackage{graphicx}
\begin{document}
\title{Raman studies of nearly half-metallic ferromagnet CoS$_2$}

\author{S.G. Lyapin}
\affiliation{Institute for High Pressure Physics of Russian Academy of Sciences, Troitsk, Russia}
\author{A.N. Utyuzh}
\affiliation{Institute for High Pressure Physics of Russian Academy of Sciences, Troitsk, Russia}
\author{A.E. Petrova}
\affiliation{Institute for High Pressure Physics of Russian Academy of Sciences, Troitsk, Russia}
\author{A.P. Novikov}
\affiliation{Institute for High Pressure Physics of Russian Academy of Sciences, Troitsk, Russia}
\author{T.A. Lograsso}
\affiliation{Ames Laboratory, Iowa State University, Ames, IA 50011, USA}
\author{S.M. Stishov}
\email{sergei@hppi.troitsk.ru}
\affiliation{Institute for High Pressure Physics of Russian Academy of Sciences, Troitsk, Russia}

\begin{abstract}
We measured the Raman spectra of ferromagnetic nearly half metal CoS$_2$ in a broad temperature range. All five Raman active modes $A_g$, $E_g$, $T_g(1)$, $T_g(2)$ and $T_g(3)$  were observed. The magnetic ordering is indicated by a change of the temperature dependences of the frequency and the line width of $A_g$ and $T_g (2) $ modes at the Curie point. The temperature dependence of the frequencies and linewidths of the $A_g$, $E_g$, $T_g(1)$, $T_g(2)$ modes in the paramagnetic phase can be described in the framework of the Klemens approach. Hardening of the $T_g(2)$, $T_g(1)$ and $A_g$ modes on cooling can be unambiguously seen in the ferromagnetic phase. The linewidths of $T_g(2)$ and $A_g$ modes behave a natural way at low  exciting laser power (decrease with decreasing temperature) in the ferromagnetic phase. At high exciting laser power the corresponding linewidths increase at temperature decreasing below the Curie temperature. Then as can be seen the line width of $A_g$ mode reaches a maxima at about 80K. This  intriging feature probably signifies a specific channel of the optical phonon decay in the ferromagnetic phase of CoS2. Tentative explanations of some of the observed effects are given, taking into account the nearly half metallic nature of CoS$_2$. 

\end{abstract}
\maketitle
\section{Introduction}
Cobalt disulphide, CoS$_2$, a metallic compound with the cubic pyrite-type crystal structure~\cite{1}, experiences a phase transition to a ferromagnetic state at $T_c\approx122$ K~\cite{2}. The magnetic and electric properties verify the itinerant nature of magnetism in CoS$_2$~\cite{2, 3, 4, 5, 6}. Upon entering the ferromagnetic state CoS$_2$ becomes a nearly half metal with a significant decrease in the density of states at the Fermi level that is reflected in an increase in the resistivity below $T_c$~\cite{7,8}. An influence of half metallicity on the thermodynamic properties and phase diagram of CoS$_2$ was reported in~\cite{9}. So one might expect that the lattice dynamic of CoS$_2$ should somehow reveal this effect as well. On the other hand the connection between the anomalous behavior of some optical modes, observed in a Raman study of the model half metal CrO$_2$, and half metallicity was not recognized~\cite{10}.  Nevertheless, because of a general interest to the half metallic  materials  we carried out a detailed Raman study of single crystals of CoS$_2$ in the temperature range 10-500 K. Raman scattering studies of CoS$_2$ at room temperature 300 K were reported before in Ref.~\cite{11,12}. We conclude that the half metallic state strongly modifies the ion-ion interaction in CoS$_2$ that therefore influences  the stretching modes in the ferromagnetic phase.    
\begin{figure}[htb]
\includegraphics[width=80mm]{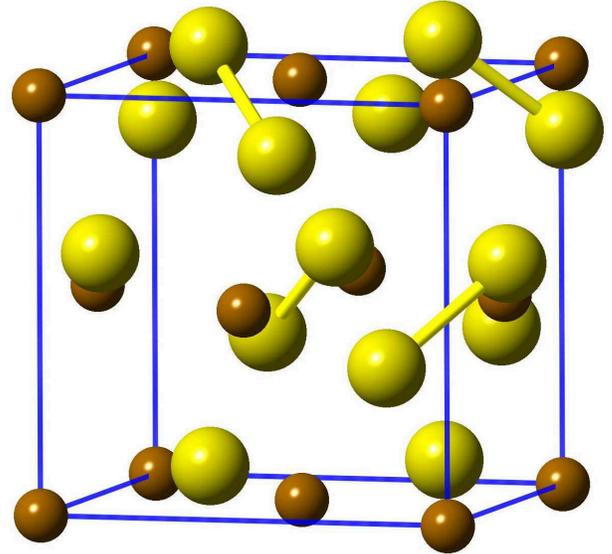}
\caption{\label{fig1} (Color online) Crystal structure of CoS$_2$. Large and small spheres stand for S and Co atoms, respectively.}
 \end{figure}

\begin{figure}[htb]
\includegraphics[width=80mm]{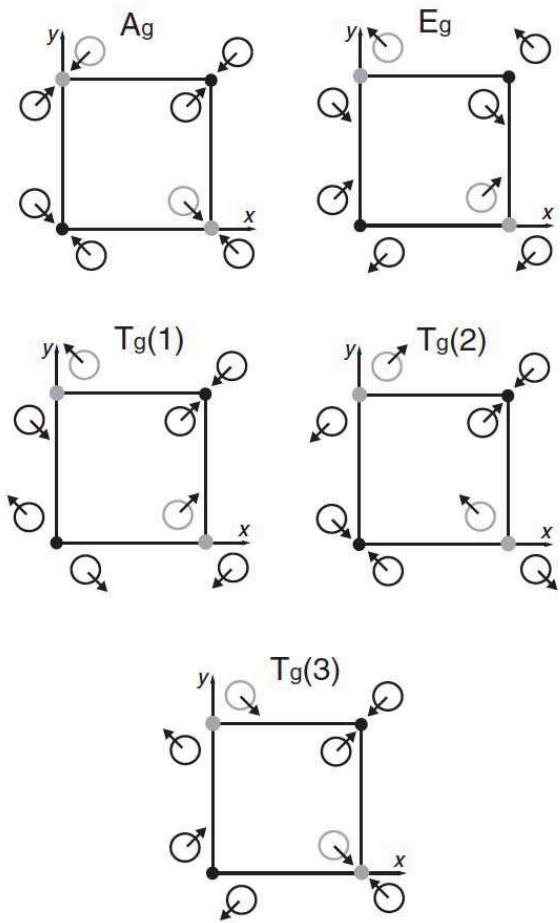}
\caption{\label{fig2} Atomic displacements for the Raman-active modes of CoS$_2$ pyrite (after ref.~\cite{14}). Open and filled circles represent S and Co atoms respectively.}
 \end{figure}
As was mentioned, the crystal structure of CoS$_2$ belongs to the pyrite type, space group Pa3, with four formula units per unit cell.  The Co atoms are situated on the sites of a face-centered cubic lattice whereas the S atoms are grouped in dumb-bell pairs oriented along [111] direction and located in the center of the cell and the midpoints of the cell edges, as is shown in Fig.~\ref{fig1}.

Five Raman modes $A_g$, $E_g$, $T_g(1)$, $T_g(2)$ and $T_g(3)$ (see Fig.~\ref{fig2}) as identified by group theory analysis~\cite{13}, can be active in the CoS$_2$ structure, and all of them were observed in the current experiments.
As is illustrated in Fig.~\ref{fig2} the $A_g$ and $T_g(2)$ modes correspond to in-phase and out-of-phase stretching vibrations of sulphur atoms in the dumbbells. $E_g$ is a pure librational mode of the dumbbells, whereas $T_g(1)$ and $T_g(3)$ modes correspond to various combinations of stretching and librational vibration.

\section{Experimental}
Raman measurements were performed using a 488 nm $Ar^+$ laser line for excitation and a triple-grating spectrometer (Princeton Instruments TriVista 555) with a liquid-nitrogen cooled CCD detector. For measurements in the temperature range 80-300K a micro-Raman setup was used: the samples were placed inside a cryostage (Linkam THMS600) and a 50x objective of a Olympus BX51 microscope was used to focus the laser and collect the Raman signal. The laser spot on the sample inside the cryostage was less than $\sim1-2\ \mu m$ in diameter. For measurements in the 10-300 K range the samples were placed in a He cryostat (Oxford Instruments OptistatSXM) and achromatic lenses were used for focusing and collecting the signal. The laser spot on the sample inside the cryostat was less than $\sim5-10\ \mu m$.  In order to avoid overheating of the samples the laser power was kept at its minimum: $\sim1$ mW, measured at the entrance window of the cryostage and $\sim7$ mW at the front window of the cryostat.

In both optical setups the true backscattering configuration was used. Polarized spectra were measured in the HH (i.e., the incident and scattering polarizations are parallel) and HV (i.e., the incident and scattering polarizations are perpendicular) configurations. The spectral slit width was less than 3 cm$^{-1}$.

Single crystals of CoS$_2$ were grown in the Ames Laboratory according to the chemical vapor transport procedure used by Wang et al.~\cite{15}. The crystals that resulted following the two-week growth run had dimensions of 1 mm or less. The crystals have a cubic pyrite-type structure with a lattice parameter a=5.5362$\pm$0.0002 ${\AA}$. The crystal quality is characterized by their high residual resistivity ratio RRR=$\rho_{297K}/\rho_{2K}=285$, which is the highest reported so far for CoS$_2$~\cite{16}.

The measurements were performed making use of three samples of CoS$_2$ thereafter designated as samples A, B, and C.

\section{Results}
\begin{figure}[htb]
\includegraphics[width=80mm]{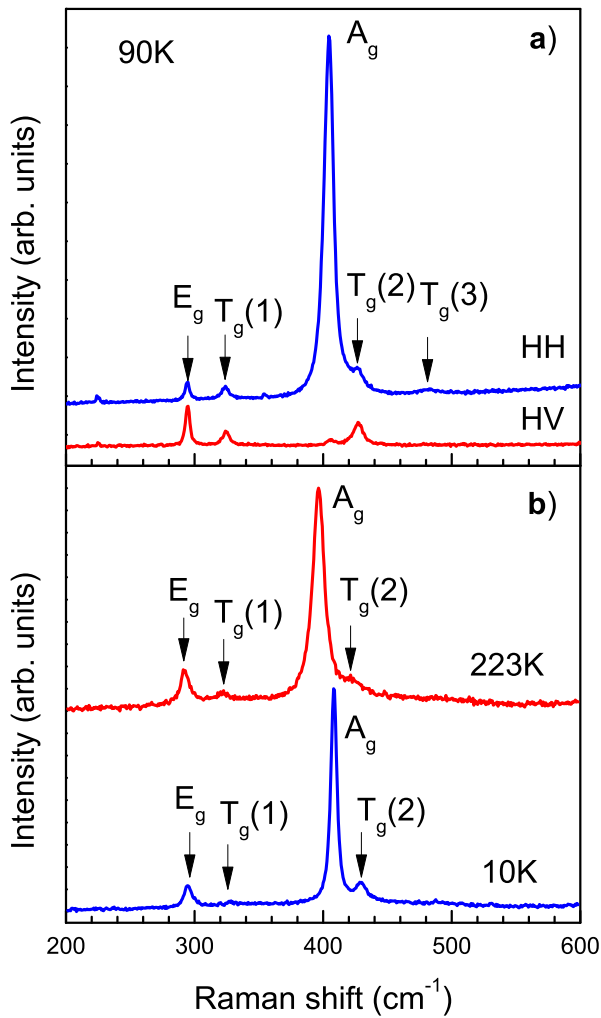}
\caption{\label{fig3} (Color online) Phonon mode assignment in the representative Raman spectra of CoS$_2$: a) sample A at 90 K in the cryostage, configurations HH (i.e., the incident and scattering radiation are parallel) and HV (i.e., the incident and scattering radiation are perpendicular) configurations, b) sample B at 10 K and 223 K in the He cryostat configuration.}
 \end{figure}

A general view of the Raman spectra of CoS$_2$ is presented in Fig.~\ref{fig3}a, b. To analyze the temperature evolution of the Raman modes, the corresponding peaks were approximated by a combination of the lorentzians that were fitted in each case.

\begin{figure}[htb]
\includegraphics[width=80mm]{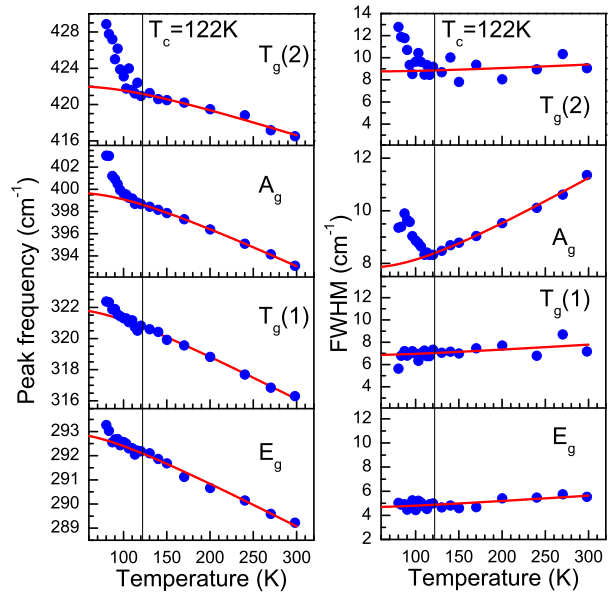}
\caption{\label{fig4} (Color online) Frequencies and  linewidths (FWHM) as a function of temperature for  $E_g$, $T_g(1)$, $A_g$ and $T_g(3)$ modes in sample A at high power (7.5 mW) laser excitation. Data for the $E_g$ and $A_g$ modes are from the HH configuration. Data for $T_g(1)$ and $T_g(3)$ modes are obtained in the HV configuration, where these modes are more pronounced. Solid lines are the least-square fitting of data taken above $T_c$=122 K  with expressions derived in the framework of the Klemens approach~\cite{17,18,19}. However, because this approach cannot be fully justified in the case of conductive materials these lines may serve as guide to the eye.}
\end{figure}

Fig.~\ref{fig4} (left panel) illustrates some features of the spectra evolution (frequencies  and linewidths), measured in geometry HH and HV in a micro-Raman cryostage. The data demonstrated in Fig.~\ref{fig4} (right panel) were obtained in a helium cryostat.

The temperature dependence of the frequencies and half-widths of the Raman modes $E_g$, $T_g(1)$, $A_g$, $T_g(2)$ using high intensity laser exitation (7.5 mW at the window of the Lincam cryostage) for sample A is shown in Fig.~\ref{fig4}.  The continuous lines in Fig.~\ref{fig4} represent results of the appoximation of the data corresponding to the paramagnetic phase of CoS$_2$ by the expressions describing the temperature dependence of the frequencies and half-widths of the Raman modes in CoS$_2$, in the spirit of the model of the decay of optical phonons developed by Klemens and others (see~\cite{17,18,19}). These "three phonon" expressions seem to be work satisfactorily in the paramagnetic phase, though one can not be sure that the Klemens approach is completely applicable in  the case of metallic crystals. The electron-phonon interaction may contribute significally to the light scattering. The "hardening" of some modes in the ferromagnetic phase obviously reflects the existence of additional channels of light scattering that  could not be discribed within the framework of the Klemens approach.

As is seen in Fig.~\ref{fig4} the measured frequencies and halfwidths of the Raman modes, possibly with exception of the frequency of $E_g$ mode and halfwidths of $T_g(1)$ and $E_g$ modes, experience some anomalies at the Curie point. The behavior of the frequencies of $A_g$, $T_g(2)$ modes in the ferromagnetic phase seem to be a natural result of mode softening in the vicinity of the phase transition. On the other hand the increase of the width of the $T_g(2)$ and $A_g$ modes with decreasing temperature appears to be counterintuitive. Note that such an effect was discovered previously in the Raman studies of CoF$_2$ and still lacks an explanation~\cite{20}. As we will see further, at low power excitation the width of the corresponding modes behaves a natural way - it decreases with decreasing temperature.

\begin{figure}[htb]
\includegraphics[width=80mm]{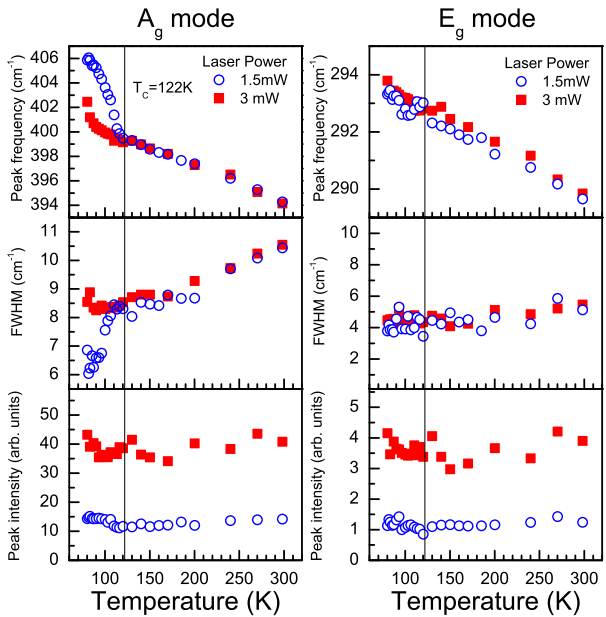}
\caption{\label{fig5} (Color online) Frequencies, linewidths and integrated intensities as a function of temperature for the $A_g$ mode (left panel) and $E_g$ mode (right panel) at low and moderate laser power excitation (micro-Raman studies of sample C).}
\end{figure}

The temperature dependence of the frequencies, widths and intensities of $A_g$ and $E_g$ modes for sample C at two different laser beam powers is displayed in Fig.~\ref{fig5}.  As follows from Fig.~\ref{fig5} the temperature dependencies of frequency, width and intensity of the $E_g$ mode, which is related to a pure librational motion of the sulphur dumbbells, do not exhibit anomalies at the Curie point at either laser power. At the same time the $A_g$ mode, corresponding to the in - phase stretching vibrations of sulphur atoms in the dumbbells, shows both a clear anomaly at the Curie point and an obvious sensitivity to the excitation power.  As is shown in Fig.~\ref{fig5} the temperature dependence of the $A_g$ frequency at low excitation power reveals a sharp turn at $\sim122$ K, indicating the existence of a well defined second order phase transition (no discontinuity is observed). But at increased laser power the phase transition features are slightly smeared out, probably due to a temperature gradient arising as a result of overheating the sample. At the same time, the width of the $A_g$ peak experiences a drastic change with increased power, as illustrated in Fig.~\ref{fig5}. The abnormal behavior of the width of the $A_g$ peak (its width increasing with decreasing temperature) is probably closely related to the factors, which define the smearing out of the phase transition. Note also that the $A_g$ and $E_g$ peak intensities are not influenced by the phase transition and temperature (Fig.~\ref{fig5}).

\begin{figure}[htb]
\includegraphics[width=80mm]{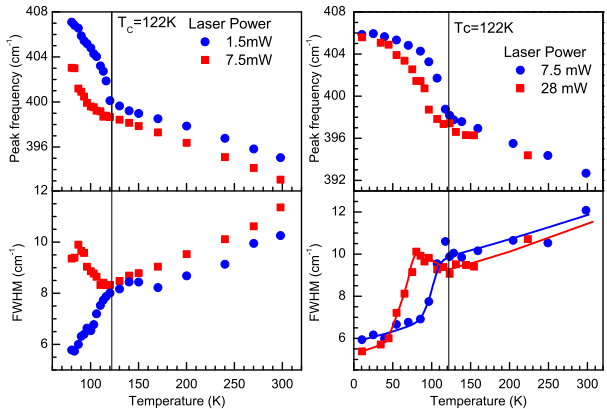}
\caption{\label{fig6} (Color online) Temperature dependence of the $A_g$ mode frequencies and linewidths  at low and high laser power excitation: left panel - results of micro-Raman studies of the sample A in the 80-300K range,  right panel - Raman studies of the sample B in the range 10-300K. Solid lines are guide to the eye.}
\end{figure}

Fig.~\ref{fig6} shows the temperature dependence of the $A_g$ mode frequencies and linewidths at low and high laser power excitation for samples A(left panel) and B (right panel). The left panel data differ little from the data in Fig.~\ref{fig5}, though they probably show that high intensity excitations decrease the frequencies of the $A_g$ mode in the paramagnetic phase. In the right panel of Fig.~\ref{fig6} the data obtained in the extended temperature range in a helium cryostat are shown. These data unambiguously demonstrate that the $A_g$ linewidth increase observed in the ferromagnetic phase does not continue this way to the lowest temperatures but reaches its maxima at $\sim80$ K. This non trivial feature obviously connected with a mechanism of the optical phonon decay in the ferromagnetic phase of CoS$_2$. We will briefly discuss this matter in the following section.  

\begin{figure}[htb]
\includegraphics[width=80mm]{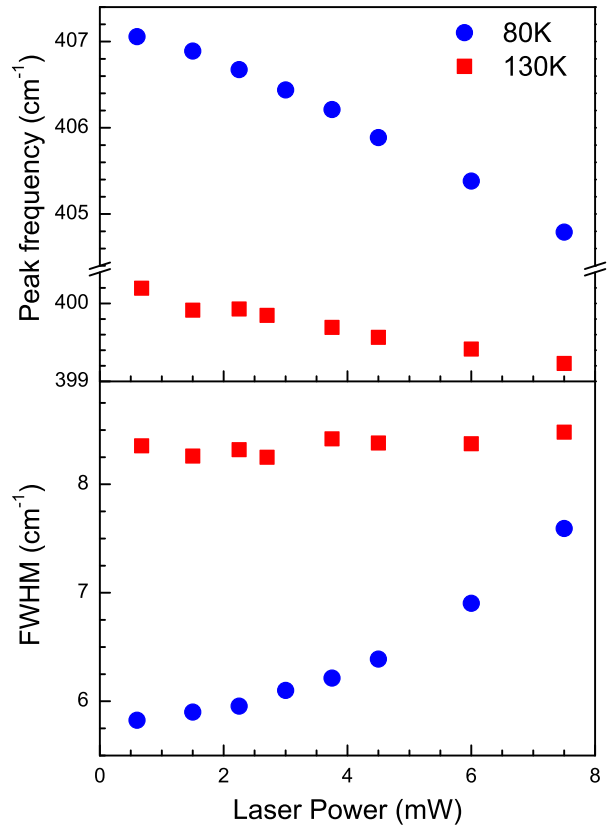}
\caption{\label{fig7} (Color online) Frequencies and linewidths as a function of the exciting laser power for the $A_g$ mode at temperature 80K (below $T_c$) and 130K (just above $T_c$), in sample A.}
 \end{figure}

Fig.~\ref{fig7} illustrates that the frequency and the linewidth of the $A_g$ mode in the ferromagnetic phase are more sensitive to the variation of the excitation power than the same quantities in the paramagnetic phase, which obviously results from the magnetic order response to the sample heating.

\section{Discussion}
Before proceeding to the discussion let's shortly summarize the main findings of the current paper.

All five Raman active modes $A_g$, $E_g$, $T_g(1)$, $T_g(2)$, and $T_g(3)$ (see Fig.~\ref{fig3}) were observed in single crystals of the itinerant ferromagnet CoS$_2$. The temperature dependence of the frequencies and linewidths of the $A_g$, $E_g$, $T_g(1)$, $T_g(2)$, modes in the paramagnetic phase can be described in the framework of the Klemens approach~\cite{17,18,19} (Fig.~\ref{fig4}).  In the ferromagnetic phase, the $T_g(2)$, $A_g$ and $T_g(1)$ modes reveal distinct hardening on cooling as is seen in Fig.~\ref{fig4}. The behavior of the $A_g$ mode indicates a smearing out of the phase transition at high excitation power (Fig.~\ref{fig5}) that possibly arises from a temperature gradient in the sample. The linewidths of $T_g(2)$ and $A_g$ modes behave in an expected way at low laser excitation power (decrease with decreasing temperature) in the ferromagnetic phase (Fig.~\ref{fig5},~\ref{fig6}). But at high laser excitation power the corresponding linewidths increase with a decrease in temperature below $T_c$.  But then, as can be seen in Fig.~\ref{fig6},  the line width of the $A_g$ mode reaches  a maximum at about 80K. At this point it should be mentioned that a number of studies of spin ordering effects on the lattice dynamics in various substances always indicate strong changes of the phonon frequency and linewidth of all or selected modes with variation of temperature (the hardening of optical phonons on cooling is common). As examples we can refer to the Raman and some IR studies of ferromagnetic semiconductors CdCr$_2$Se$_4$ and CdCr$_2$S$_4$~\cite{21}, ferromagnetic metal La$_0.7$Ca$_0.3$MnO$_3$~\cite{22}, strongly correlated ferromagnetic metal SrRuO$_3$~\cite{23}, half metal CrO$_2$~\cite{10}, antiferromagnetic halides of transition metals~\cite{24}, antiferromagnetic metal $\alpha$-MnSe~\cite{25}, etc..  As seen in the cited references quite different explanations were suggested to explain the observed effects. So far no general mechanism was identified. As for the case of CoS$_2$ L.Falkovsky in a recent paper~\cite{26} interpreted some of our data on the Raman scattering in CoS$_2$ more or less successfully in the terms of the phonon-electron interaction, arising from interband transitions at finite temperature. As Falkovsky's analysis shows the expected direct phonon-magnon interactions are unable to contribute significantly to the observed effects. Note that according to the theory~\cite{26} the phonon-electron interactions in CoS$_2$ influence the line shift and width both in the ferromagnetic and paramagnetic phases.
\begin{figure}[htb]
\includegraphics[width=80mm]{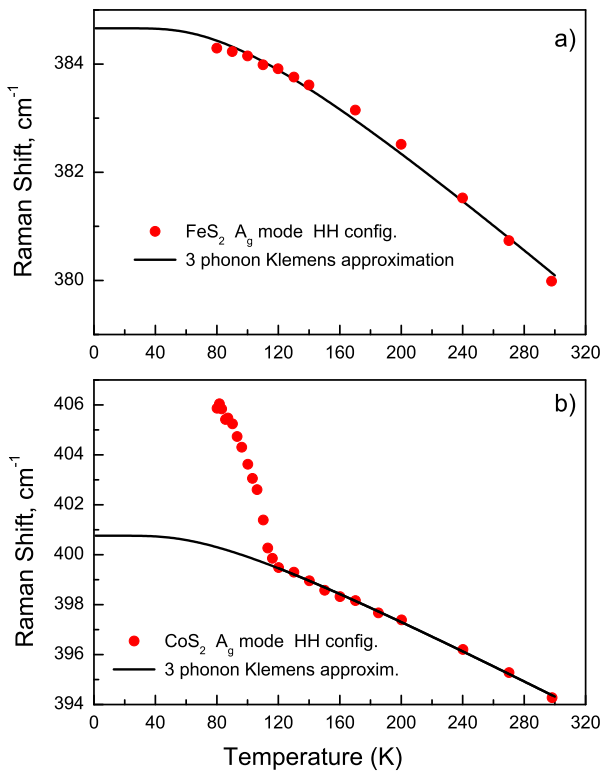}
\caption{\label{fig8} (Color online) Frequency shift of $A_g$ mode in FeS$_2$ (a) and CoS$_2$ (b)(FeS$_2$ data were obtained in the course of the current study).}
 \end{figure}

Our view of the Raman scattering features in CoS$_2$ is somewhat different. The nearly half metal nature of the ferromagnetic phase of CoS$_2$ creates a specific situation due to a spin-dependent band structure effect.  This strongly modifies the ion-ion interaction in ferromagnetic CoS$_2$ as is evidenced by a decrease of its compressibility~\cite{9}. Correspondingly one may expect higher frequencies of stretching modes in the ferromagnetic phase CoS$_2$ compared to the paramagnetic phase (the hardening effect).  The puzzling maximum in the line width, which appears below the Curie temperature at high laser power excitation (see Figs.~\ref{fig5},~\ref{fig6}), most probably resulted from specifics of the decay mechanism in the ferromagnetic phase of CoS$_2$, which certainly includes the Klemence anharmonic contribution. On the other hand the optical phonons, as well as the thermal excitations,  could directly or indirectly turn over the spins in the majority band, therefore filling out the minority band in CoS$_2$. This action defines another channel  of the decay in the ferromagnetic phase. The maximum in the line width probably implies that there exists one more  channel of the decay, which reveals itself at high laser power excitation and which is almost exhausted at temperatures around 80 K. \section{Conclusion}
In conclusion,we measured the Raman spectra of ferromagnetic nearly half metal CoS$_2$ in a broad temperature range. The magnetic ordering is signified by a change of the temperature dependences of the frequency and the line width of $A_g$ and $T_g (2)$ modes at the Curie point. The Raman intensities are not practically
affected by the magnetic ordering. Variations of frequencies of the Raman stretching modes in the ferromagnetic phase of CoS$_2$ define solely by the modified ion-ion interaction,  which is altered due to  the nearly half metallic character of the phase. This situation ceases to exist above the Curie point, and it is plausible that the Klemens anharmonic approach is valid in the paramagnetic phase. The latter seems to be supported by the experiment (Fig.~\ref{fig8}). In Fig.~\ref{fig8} the temperature dependencies of the $A_g$  modes of CoS$_2$ and FeS$_2$ are compared. As is seen the Klemens approach works satisfactorily for FeS$_2$ and for the paramagnetic phase CoS$_2$, which confirms our supposition. In the end, the maximum in the line widths  below the Curie temperature probably implies that there exists a specific  channel of the decay, which reveal itself at high laser power excitation. 
\section{Acknowledgements}
This work was supported by the Russian Foundation for Basic Research (grant 12-02-00376), Program of the Physics Department of RAS on Strongly Correlated Electron Systems and Program of the Presidium of RAS on Strongly Compressed Matter.

\end{document}